\documentclass[sigconf]{acmart}

\usepackage{booktabs} 
\usepackage{paralist}
\usepackage{algorithm}
\usepackage{subfigure}
\usepackage[strings]{underscore}

\copyrightyear{2018} 
\acmYear{2018} 
\setcopyright{acmcopyright}
\acmConference[HT'18]{29th ACM Conference on Hypertext and Social Media}{July 9--12, 2018}{Baltimore, MD, USA}
\acmBooktitle{HT'18: 29th ACM Conference on Hypertext \& Social Media, July 9--12, 2018, Baltimore, MD, USA}
\acmPrice{15.00}
\acmDOI{10.1145/3209542.3209555}
\acmISBN{978-1-4503-5427-1/18/07} 

\begin{document}

\title{Search Rank Fraud De-Anonymization in Online Systems}

\author{Mizanur Rahman}
\affiliation{
\institution{Florida Int'l University, USA}
}
\email{mrahm031@fiu.edu}
\author{Nestor Hernandez}
\affiliation{
\institution{Florida Int'l University, USA}
}
\email{nestorghh@gmail.com}
\author{Bogdan Carbunar}
\affiliation{
\institution{Florida Int'l University, USA}
}
\email{carbunar@gmail.com}
\author{Duen Horng Chau}
\affiliation{
\institution{Georgia Tech}
}
\email{polo@gatech.edu}

\begin{abstract}
We introduce the {\it fraud de-anonymization} problem, that goes beyond fraud detection, to unmask the human masterminds responsible for posting search rank fraud in online systems. We collect and study search rank fraud data from Upwork, and survey the capabilities and behaviors of 58 search rank fraudsters recruited from 6 crowdsourcing sites. We propose Dolos, a fraud de-anonymization system that leverages traits and behaviors extracted from these studies, to attribute detected fraud to crowdsourcing site fraudsters, thus to real identities and bank accounts. We introduce MCDense, a min-cut dense component detection algorithm to uncover groups of user accounts controlled by {\it different} fraudsters, and leverage stylometry and deep learning to attribute them to crowdsourcing site profiles. Dolos correctly identified the owners of 95\% of fraudster-controlled communities, and uncovered fraudsters who promoted as many as 97.5\% of fraud apps we collected from Google Play. When evaluated on 13,087 apps (820,760 reviews), which we monitored over more than 6 months, Dolos identified 1,056 apps with suspicious reviewer groups. We report orthogonal evidence of their fraud, including fraud duplicates and fraud re-posts.
\end{abstract}


\maketitle

\section{Introduction}

\begin{figure}
\centering
\includegraphics[width=0.47\textwidth]{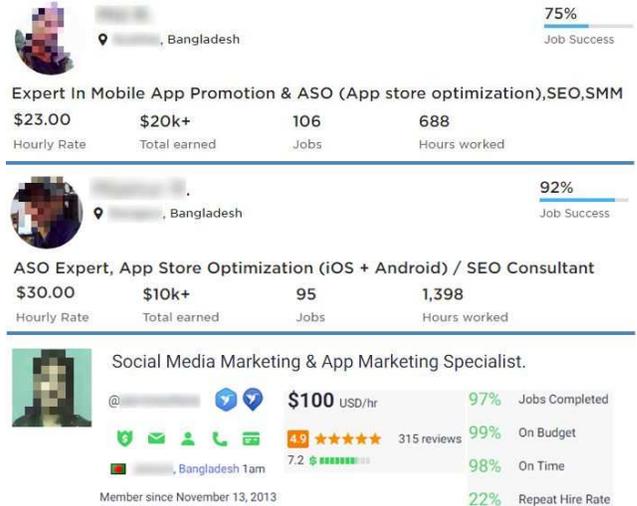}
\vspace{-10pt}
\caption{Anonymized snapshots of profiles of search rank fraudsters from Upwork (top 2) and Freelancer (bottom). Fraudsters control hundreds of user accounts and earn thousands of dollars through hundreds of work hours. Our goal is to de-anonymize fraud, i.e., attribute fraud detected for products in online systems, to the crowdsourcing site accounts of the fraudsters (such as these) who posted it.}
\label{fig:experts}
\vspace{-15pt}
\end{figure}

The competitive, dynamic nature of online services provides high rewards to the developers of top ranking products, through direct payments or ads. The pressure to succeed, coupled with the knowledge that statistics over user actions (e.g., reviews, likes, followers, app installs) play an essential part in a product's ranking~\cite{HP,L11,Ankit.Jain}, has created a black market for {\it search rank fraud}: Fraudsters create hundreds of user accounts, connect with product developers through crowdsourcing sites~\cite{Fiverr,Upwork,Freelancer}, then post fake activities for their products, from the accounts they control, see Figure~\ref{fig:experts}. 


Detecting and disincentivizing search rank fraudsters are tasks of paramount importance to building trust in online services and the products that they host. Previous work has focused mainly on detecting online fraud~\cite{CHZY17,RRCC16,MLG12,WGF17,fei2013exploiting,mukherjee2013spotting,li2017bimodal,hooi2016birdnest,badri2016uncovering,ACF13,hooi2016fraudar,rayana2015collective,akoglu2013opinion,XZ14}, and many review based online systems filter out detected fraudulent activities~\cite{Google.filter,Amazon.filter,MVLG13}. However, a preliminary study we performed with 58 fraudsters from 6 crowdsourcing sites revealed that workers with years of search rank fraud expertise are actively working on such jobs, and are able to post hundreds of reviews for a single product at prices ranging from a few cents to \$10 per review, see e.g., Figure~\ref{fig:experts:survey}. This suggests that fraud detection alone is unable to prevent large scale search rank fraud behaviors in online systems.

In this paper we introduce the {\it fraud de-anonymization} problem, a new approach to address the limitations of status quo solutions, through disincentivizing search rank fraud workers and their employers. Unlike standard de-anonymization, which refers to the adversarial process of identifying users from data where their Personally Identifiable Information (PII) has been removed, the fraud de-anonymization problem seeks to attribute detected search rank fraud to the humans who posted it. A solution to this problem will enable online services to put a face to the fraud posted for the products they host,
retrieve banking information and use it to pursue fraudsters, and provide proof of fraud to customers, e.g., links to the crowdsourcing accounts responsible, see Figure~\ref{fig:experts}. Thus, fraud de-anonymization may provide counter-incentives both for crowdsourcing workers to participate in fraud jobs, and for product developers to recruit fraudsters.

To understand and model search rank fraud behaviors, we have developed a questionnaire and used it to survey 58 fraudsters recruited from 6 crowdsourcing sites. We have collected data from search rank fraud jobs and worker accounts in Upwork, and used it to identify fraudster traits and to collect 111,714 fake reviews authored by 2,664 fraudulent Google Play accounts, controlled by an \textit{expert core} among 533 identified search rank fraudsters.

We leverage the identified traits to introduce {\scshape Dolos}\footnote{{\scshape Dolos} is a concrete block used to protect harbor walls from erosive ocean waves.}
a system that cracks down fraud by unmasking the human masterminds responsible for posting significant fraud.
{\scshape Dolos} detects then attributes fraudulent user accounts in the online service, to the crowdsourcing site accounts of the workers who control them. We devise MCDense, a min-cut dense component detection algorithm that analyzes common activity relationships between user accounts to uncover groups of accounts, each group controlled by a different search rank fraudster. We further leverage stylometry, graph based deep learning feature extraction tools, and supervised learning to attribute MCDense detected groups to the crowdsourcing fraudsters who control them.

{\scshape Dolos} correctly attributed 95\% of the reviews of 640 apps (that received significant, ground truth search rank fraud) to their authors.  For 97.5\% of the apps, {\scshape Dolos} correctly de-anonymized at least one of the fraudsters who authored their fake reviews. {\scshape Dolos} achieved 90\% precision and 89\% recall when attributing the above 2,664 fraudulent accounts to the fraudsters who control them. Further, MCDense significantly outperformed an adapted densest subgraph solution.

We have evaluated {\scshape Dolos} on 13,087 Google Play apps (and their 820,760 reviews) that we monitored over more than 6 months. {\scshape Dolos} discovered that 1,056 of these apps have suspicious reviewer groups. Upon close inspection we found that (1) 29.9\% of their reviews were {\it duplicates} and (2) 73\% of the apps that had at least one MCDense discovered clique, received reviews from the expert core fraudsters that we mentioned above. We also report cases of {\it fraud re-posters}, accounts who re-post their reviews, hours to days after Google Play filters them out (up to 37 times in one case). In summary, we introduce the following contributions:


\begin{compactitem}

\item
{\bf Fraud de-anonymization problem formulation}. Introduce a new approach to combat and disincentivise search rank fraud in online systems.

\item
{\bf Study and model search rank fraud}. Survey 58 fraudsters from 6 crowdsourcing websites, collect gold standard attributed search rank fraud data and extract insights into fraudster behaviors.

\item
{\bf {\scshape Dolos} and MCDense}. Exploit extracted insights to develop fraud de-anonymization algorithms. Evaluate algorithms extensively on Google Play data. Identify orthogonal evidence of fraud from detected suspicious products. The code is available for
download at \url{https://github.com/FraudHunt}.

\end{compactitem}

\section{Study \& Model Search Rank Fraud}
\label{sec:model}

\subsection{System and Adversary Model}

\begin{figure}
\centering
\includegraphics[width=3.19in]{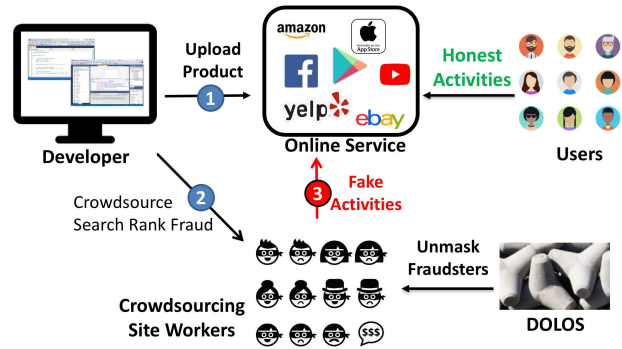}
\vspace{-10pt}
\caption{ {\bf System and adversary model}. Developers upload products, on which users post activities, e.g., reviews, likes. Adversarial developers crowdsource search rank fraud. Unlike fraud detection solutions, {\bf {\scshape Dolos} unmasks the human fraudsters responsible for posting search rank fraud}. 
\label{fig:system:model}}
\vspace{-15pt}
\end{figure}

We consider an ecosystem that consists of online services and crowdsourcing sites. Online services host accounts for developers, products and users, see Figure~\ref{fig:system:model}. Developers use their accounts to upload products. Users post {\it activities} for products, e.g., reviews, ratings, likes, installs.
%
%
Product accounts display these activities posted and statistics, while user accounts list the products on which users posted activities.
%
%
Crowdsourcing sites host accounts for {\it workers} and {\it employers}. Worker accounts have unique identifiers and bank account numbers used to deposit the money that they earn.  Employers post {\it jobs}, while workers bid on jobs, and, following negotiation steps, are assigned
or win the jobs.

\begin{figure*}[t]
\centering
\includegraphics[width=0.97\textwidth]{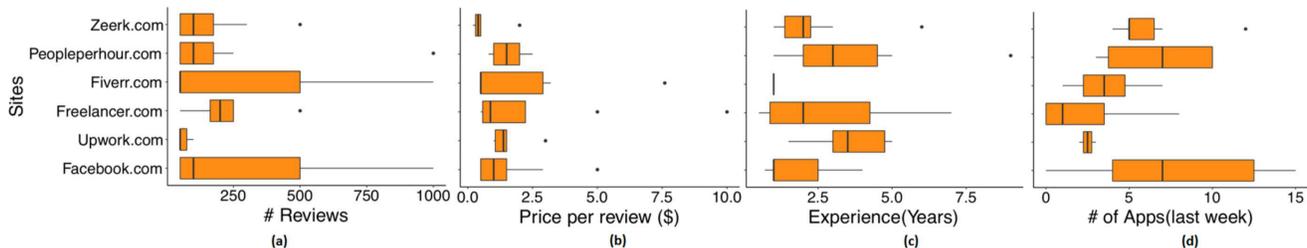}
\vspace{-15pt}
\caption{Statistics over 44 fraudsters (targeting Google Play apps) recruited from 5 crowdsourcing sites: minimum, average and maximum for
(a) number of reviews that a fraudster can write for an app,
(b) price demanded per review,
(c) years of experience,
(d) number of apps reviewed in the past 7 days. Fraudsters report to be able to write hundreds of reviews for a single app, have years of experience and are currently active. Prices range from 56 cents to \$10 per review.}
\label{fig:experts:survey}
\vspace{-10pt}
\end{figure*}


We consider product developers who hire workers from crowdsourcing sites, to perform search rank fraud, see Figure~\ref{fig:experts}. 
We focus on workers who control multiple user accounts in the online system, which they use to post fake activities, e.g., review, rate, install.

Previous studies of online fraud include the work of Yang et al.~\cite{yang2012analyzing}, who showed that ``criminal'' Twitter accounts tend to form small-world social networks. Mukherjee et al.~\cite{MLG12,mukherjee2013spotting} confirmed this finding and introduced features that identify reviewer groups, who review many products in common but not much else, post their reviews within small time windows, and are among the first to review the product. Further, Beutel et al.~\cite{BXGPF13} proposed CopyCatch, a system that identifies {\it lockstep behaviors}, i.e., groups of user accounts that act in a quasi-synchronized manner, to detect fake page likes in Facebook. Chen et al.~\cite{CHZY17} identify clusters of apps in Apple's China App store, that have been promoted in a similar fashion.

In contrast, in this paper we focus on de-anonymizing fraud, by attributing it to fraudsters recruited from crowdsourcing sites. In this section we describe our efforts to understand and model such search rank fraud experts.
In the following, we use the terms {\it worker}, {\it fraudster} and {\it fraud worker}, interchangeably.

\subsection{Motivation: Fraudster Capabilities}


To evaluate the magnitude of the problem, we have first contacted 44 workers from several crowdsourcing sites including Zeerk (12), Peopleperhour (9), Freelancer (8), Upwork (6) and Facebook groups (9), who advertised search rank fraud capabilities for app markets. We asked them (1) how many reviews they can write for one app, (2) how much they charge for one review, (3) how many apps they reviewed in the past 7 days, and (4) for how long they been active in promoting apps.

Figure~\ref{fig:experts:survey} shows statistics over the answers, organized by crowdsourcing site. It suggests significant profits for fraudsters, who claim to be able to write hundreds of reviews per app (e.g., an average of 250 reviews by Freelancer workers) and charge from a few cents (\$0.56 on average from Zeerk.com workers) to \$10 per review (Freelancer.com). Fraudsters have varied degrees of expertise in terms of years of experience and recent participation in fraud jobs. For instance, fraudsters from Peopleperhour and Upwork have more than 2.5 years experience and more than 3 recent jobs on average. Further, in recently emerged Facebook groups, that either directly sell reviews or exchange reviews,
fraudsters have less than 2.5 years experience, but are very active, with more than 7 jobs in the past 7 days on average, and economical (\$1.3 on average per review).

Subsequently, we have developed a more detailed questionnaire to better understand search rank fraud behaviors and delivered it to 14 fraud freelancers that we recruited from Fiverr. 
We paid each participant \$10, for a job that takes approx. 10 minutes. The IPs from which the questionnaire was accessed revealed that the participants were from Bangladesh (5), USA (2), Egypt (2), Netherlands, UK, Pakistan, India and Germany (1). The participants declared to be male, 18 - 28 years old, with diverse education levels: less than high school (1), high school (2), associate degree (3), in college (5), bachelor degree or more (3).

The participants admitted an array of fraud expertise (fake reviews and ratings in Google Play, iTunes, Amazon, Facebook and Twitter, fake installs in Google Play and iTunes, fake likes and followers in Facebook and Instagram, influential tweets in Twitter). We found a mix of (1) inexperienced and experienced fraudsters: 4 out of 14 had been active less than 2 months and 6 fraudsters had been active for more than 1 year, and (2) active and inactive fraudsters: 4 had not worked in the past month, 9 had worked on 1-5 fraud jobs in the past month, and 1 worked on more than 10 jobs; 8 fraudsters were currently active on 1-5 fraud jobs, and 1 on more than 5. Further, we observed varying search rank fraud capabilities: 8 of the 11 surveyed fraudsters who wrote reviews, admit to have reviewed an app at least 5 times; 1 admits to have written 51 to 100 reviews for an app.

Of the 14 fraudsters surveyed, 3 admitted to working in teams that had more than 10 members, and to sharing the user accounts that they control, with others. 10 fraudsters said that they control more than 5 Google Play accounts and 1 fraudster had more than 100 accounts. Later in this section we show that this is realistic, as other 23 fraudsters we recruited, were able to reveal between 22 and 86 Google Play accounts that they control. Further, 4 fraudsters said that they never abandon an account, 5 said that they use each account until they are unable to login, and 4 said that they use it for at most 1 year. This is confirmed by our empirical observation of the persistence of fraud (see end of section \ref{sec:model:fpc}).



%

\noindent
{\bf Ethical considerations}.
We have developed our protocols to interact with participants and collect data in an IRB-approved manner (Approval \#: IRB-15-0219@FIU).


\subsection{A Study of Search Rank Fraud Jobs}

We identified and collected data from 161 search rank fraud jobs in Upwork that request workers to post reviews on, or install Google Play and iTunes apps.  We have collected the $533$ workers who have bid on these jobs.  We call the bidding workers that are awarded a job, {\it winners}. One job of the 161, was awarded to 12 workers; more jobs were awarded to 2 workers than to only 1. This indicates that hiring multiple workers is considered beneficial by adversarial developers, and suggests the need to attribute detected organized fraud activities to human masterminds (see next section).

We introduce the concepts of {\it co-bid} and {\it co-win graphs}. In the co-bid graph, nodes are workers who bid on fraud jobs; edges connect workers who bid together on at least one job. The edge weights denote the number of jobs on which the endpoint workers have bid together. In the co-win graph, the weight of an edge is the number of fraud jobs won by both endpoint workers.

\begin{figure}
\centering
\vspace{-12pt}
\subfigure[]
{\label{fig:co:bidder}{\includegraphics[width=1.5in]{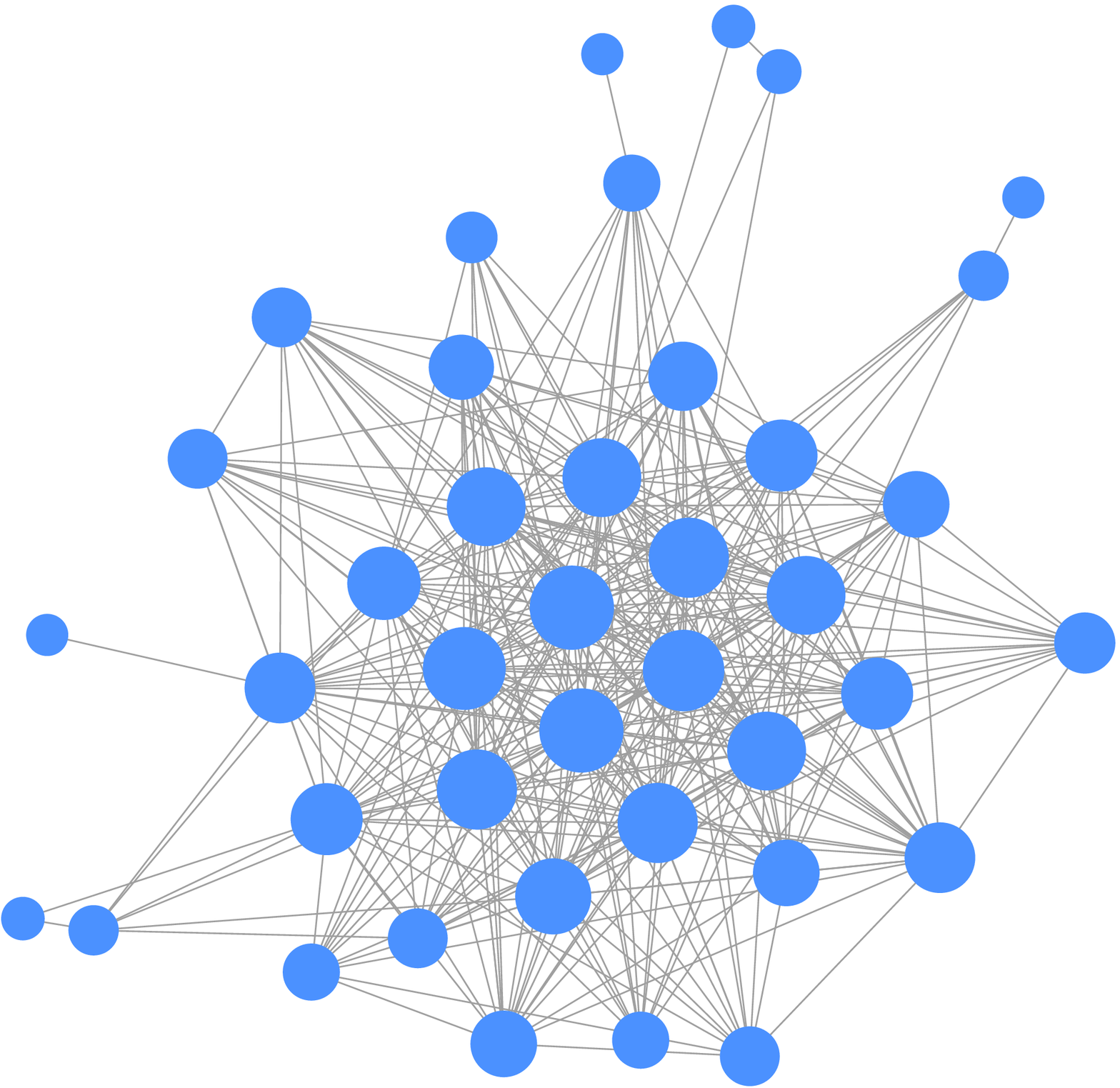}}}
\subfigure[]
{\label{fig:co:winner}{\includegraphics[width=1.25in]{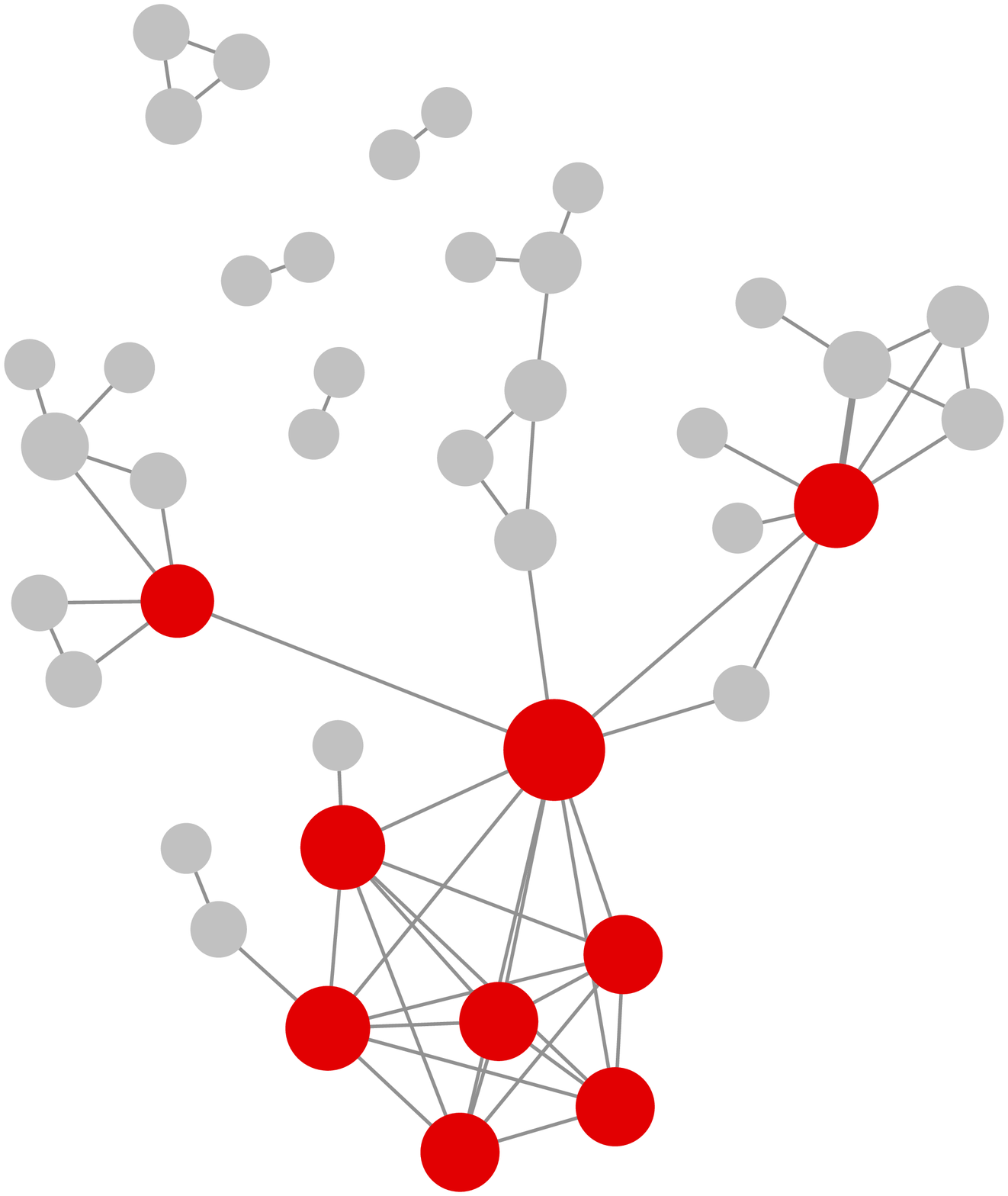}}}
\vspace{-15pt}
\caption{(a) \textbf{Worker co-bid graph}: Nodes are Upwork workers.  An edge connects two workers who co-bid on search rank fraud jobs. We see a tight co-bid community of workers; some co-bid on 37 jobs.
(b) \textbf{Worker Co-win graph} with an ``expert core'' of $8$ workers (red), each winning $8-15$ jobs. Edges connect workers who won at least one job together. Any two workers collaborated infrequently, up to 4 jobs.
}
\vspace{-20pt}
\end{figure}

Out of the 56 workers who won the 161 jobs, only 40 had won a job along with another bidder. Figure~\ref{fig:co:bidder} shows the co-bid graph of these 40 winners, who form a tight community. Figure~\ref{fig:co:winner} plots the co-win graph of the 40 winners. We observe an ``expert core'' of 8 workers who each won between 8 to 15 jobs. Further, we observe infrequent collaborations between any pair of workers: any two workers collaborated on at most 4 jobs.

\noindent
{\bf Empirical Adversary Traits}.
Our studies reveal several search rank fraudster traits:

\begin{compactitem}

\item
{\bf Trait 1}:
Fraudsters control multiple user accounts which they use to perpetrate search rank fraud.

\item
{\bf Trait 2}:
While fraudsters have diverse search rank fraud capabilities, crowdsourcing sites have an ``expert core'' of successful search rank fraud workers. Many fraudsters are willing to contribute, but few have the expertise or reputation to win such jobs.

\item
{\bf Trait 3}:
Search rank fraud jobs often recruit multiple workers. Thus, targeted products may receive fake reviews from multiple fraudsters.

\item
{\bf Trait 4}:
Any two fraudsters collaborate infrequently, when compared to the number of search rank fraud jobs on which they have participated, see Figure~\ref{fig:co:winner}.


\item
{\bf Trait 5}:
Fraudsters, including experts, are willing to share information about their behaviors, perhaps to convince prospective employers of their expertise.

\end{compactitem}

\noindent
{\scshape Dolos} exploits these traits to detect and attribute groups of fraudulent user accounts to the fraudsters who control them. While we do not claim that the sample data from which the traits are extracted is representative, in the evaluation section we show that {\scshape Dolos} can accurately de-anonymize fraudsters.

\subsection{Fraudster Profile Collection (FPC)}
\label{sec:model:fpc}

\begin{figure}
\centering
\includegraphics[width=0.49\textwidth]{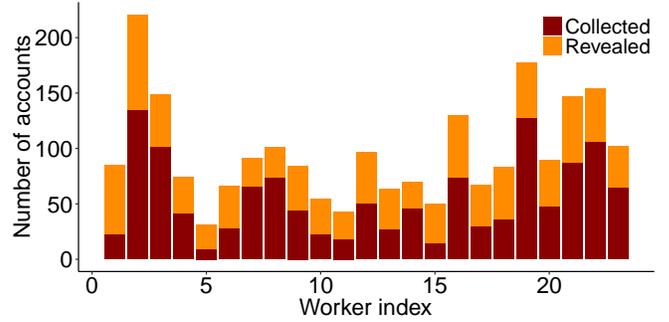}
\vspace{-20pt}
\caption{\textbf{Attributed, fraudster-controlled accounts.} The numbers of Google Play accounts revealed by the detected fraudsters are shown in red. Each of the 23 fraudsters has revealed between 22 to 86 accounts. Guilt-by-association accounts are shown in orange. We have collected a total of 2,664 accounts (red + orange). One fraudster controls (at least) 217 accounts.}
\label{fig:worker:accounts}
\vspace{-15pt}
\end{figure}

Kaghazgaran et al.~\cite{KCA17} identified crowdsourcing site jobs that reveal the targeted Amazon products, then studied those products. However, they did not attribute the fraudulent reviews to the crowdsourcing site accounts of the workers who worked on those jobs. Xie and Zhu~\cite{XZ15} monitored 52 paid review service providers for four months and exposed apps that they promoted. 

Unlike previous work, we leveraged Trait 5 to collect a first gold standard dataset of attributed, fraudster controlled accounts in Google Play. For this, we have identified and contacted 100 Upwork, Fiverr and Freelancer workers with significant bidding activity on search rank fraud jobs targeting Google Play apps.
Figure~\ref{fig:worker:accounts} shows the number of accounts (bottom, red segments) revealed by each of 23 most responsive of these workers: between 22 and 86 Google Play accounts revealed per worker, for a total of 1,356 user accounts.

\noindent
{\bf Fraud app dataset}.
To expand this data, we collected first a subset of 640 apps that received the highest ratio of reviews from accounts controlled by the above 23 expert core workers to the total number of reviews. We have monitored the apps over a 6 month interval, collecting their new reviews once every 2 days. The 640 apps had between 7 to 3,889 reviews. Half of these apps had at least 51\% of their reviews written from accounts controlled by the 23 fraudsters. In the following we refer to these, as the {\it fraud apps}.

\noindent
{\bf Union fraud graph}.
We have collected the account data of the $38,123$ unique reviewers ($956$ of which are the seed accounts revealed by the 23 fraudsters)  of the fraud apps, enabling us to build their {\it union fraud graph}: a node corresponds to an account that reviewed one of these apps (including fraudster controlled and honest ones), and the weight of an edge denotes the number of apps reviewed in common by the accounts that correspond to the end nodes.  We have removed duplicates: an account that reviewed multiple fraud apps has only one node in the graph. The union fraud graph has 19,375,550 edges and 162 disconnected components, of which the largest has 37,566 nodes.

\begin{table}
\setlength{\tabcolsep}{.16677em}
\centering
\small
\textsf{
\begin{tabular}{l | r | r | r }
\toprule
\textbf{Algorithm } & \textbf{ Precision } & \textbf{ Recall } & \textbf{ F-measure}\\
\midrule
RF & 95.5\% & 91.6\% & 93.5\%\\
\textbf{SVM} & \textbf{98.5\%} & \textbf{98.3\%} & \textbf{98.5\%}\\
k-NN & 97.1\% & 96.4\% & 96.7\%\\
MLP & 98.6\% & 98.1\% & 98.4\%\\
\bottomrule
\end{tabular}
}
\caption{Account attribution performance on gold standard fraudster-controlled dataset, with several supervised learning algorithms
(parameters $d=300$, $t=100$, $\gamma=80$, and $w=5$ set through a grid search). SVM performed best.}
\label{table:deepwalk:piper}
\vspace{-25pt}
\end{table}

\noindent
{\bf Guilt-by-association}.
We have labeled each node of the union fraud graph with the ID of the fraudster controlling it or with ``unknown'' if no such information exists. For each unknown labeled node $U$, we decide if $U$ is controlled by one of the fraudsters, based on how well $U$ is associated with accounts controlled by the fraudster. However, $U$ may be connected to the accounts of multiple fraudsters (see Trait 3).

To address this problem, we leveraged Trait 4 to observe that random walks that start from nodes controlled by the same fraudsters are likely to share significant context, likely different from the context of nodes controlled by other fraudsters, or that are honest. We have pre-processed the union fraud graph to convert it into a non-weighted graph: replace an edge between nodes $u_i$ and $u_j$ with weight $w_{ij}$, by $w_{ij}$ non-weighted edges between $u_i$ and $u_j$. We then used the DeepWalk algorithm~\cite{PARS14} to perform $\gamma$ random walks starting from each node $v$ in this graph, where a walk samples uniformly from the neighbors of the last vertex visited until it reaches the maximum walk length ($t$). The pre-processing of the union graph ensures that the probability of DeepWalk at node $u_i$ to choose node $u_j$ as next hop, is proportional to $w_{ij}$. DeepWalk also takes as input a window size $w$, the number of neighbors used as the context in each iteration of its SkipGram component. Deepwalk returns a $d$-dimensional representation in $\mathbb{R}^d$ for each of the nodes. We then used this representation as predictor features for the ``ownership'' of the account $U$ - the fraudster who controls it.


Table~\ref{table:deepwalk:piper} highlights precision, recall, and F-measure achieved by different supervised learning algorithms. We observe that SVM reaches $98.5\%$ F-measure which suggests DeepWalk's ability to provide useful features and assist in our guilt-by-association process. We then applied the trained model to the remaining and unlabeled  accounts in the union fraud graph obtaining new guilt-by-association accounts for each of the 23 workers.  Figure~\ref{fig:worker:accounts} shows the number of seed and guilt-by-association accounts uncovered for each of the 23 fraudsters. We have collected $1,308$ additional accounts across workers for a total of 2,664 accounts. 

\noindent
{\bf Persistence of fraud}.
After more than 1 year following the 
collection of the 2,664 fraudster-controlled accounts, we have re-accessed the accounts. We found that 67 accounts had been deleted and 529 accounts were inactive, i.e., all information about apps installed, reviewed, +1'd was removed. 2,068 accounts were active. This is consistent with the findings from our fraudster survey, where 4 out of 14 surveyed fraudsters said that they never abandon an account, 5 said that they use each account until they are unable to login, and 4 said that they use it for at most 1 year. This further suggests the limited ability of Google Play to identify and block fraudster-controlled accounts.

\section{Fraud De-Anonymization System}

\begin{figure}[t]
\centering
\includegraphics[width=0.50\textwidth]{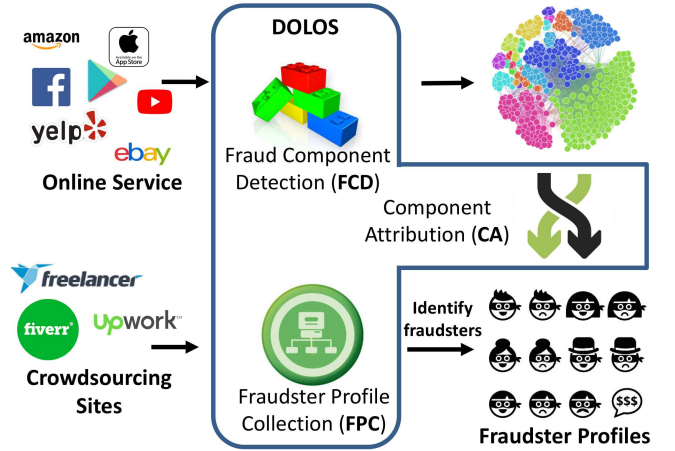}
\caption{{\bf {\scshape Dolos} system architecture}. The Fraud Component Detection (FCD) module partitions the co-activity graphs of apps into loosely inter-connected, dense components. The Component Attribution (CA) module attributes FCD detected components to fraudster profiles collected by the Fraudster Profile Collector (FPC), see $\S$~\ref{sec:model:fpc}.}
\label{fig:f2f}
\vspace{-15pt}
\end{figure}

\subsection{Problem Definition}

Unlike standard de-anonymization, which refers to the adversarial process of identifying users from data where their PII has been removed, in this paper we define the fraud de-anonymization problem in a positive context. Specifically, let $\mathcal{W} = \{ W_1, .., W_n \}$ be the set of crowdsourcing worker accounts, let $\mathcal{U} = \{ U_1, .., U_m \}$ be the set of user accounts and let $\mathcal{A} = \{ A_1, .., A_a \}$ be the set of products hosted by the online service. Then, given a product $A \in \mathcal{A}$, return the subset of fraudsters in $\mathcal{W}$ who control user accounts in $\mathcal{U}$ that posted fraudulent activities for $A$.

\subsection{Solution Overview}

We introduce {\scshape Dolos}, the first fraud de-anonymization system that integrates activities on both crowdsourcing sites and online services. As illustrated in Figure~\ref{fig:f2f}, {\scshape Dolos} (1) proactively identifies new fraudsters and builds their profiles in crowdsourcing sites, then (2) processes product and user accounts in online systems to attribute detected fraud to these profiles. The gold standard fraudster profile collection (FPC) module described in the previous section performs the first task. In the following, we focus on the second task, which we break into two sub-problems:

\begin{compactitem}

\item
{\bf Fraud-Component Detection Problem}.
Given a product $A \in \mathcal{A}$, return a set of components $C_A = \{ C_1, .., C_k \}$, where any $C_{j = 1 .. k}$ consists of a subset of the user accounts who posted an activity for $A$, s.t., those accounts are either controlled by a single worker in $\mathcal{W}$, or are honest.

\item
{\bf Component Attribution Problem}.
Given $\mathcal{W}$ and a component $C \in C_A$, return the identity of the worker in $\mathcal{W}$ who controls all the accounts in the component, or $\perp$ if the accounts are not controlled by a worker.

\end{compactitem}

\noindent
The FCD module of {\scshape Dolos} partitions the reviews of a product into components, such that all the reviews in a component were posted by a single fraudster. The CA module attributes each component to crowdsourcing account of the fraudster who controls it.
%
In the following, we detail these modules.


\subsection{Fraud Component Detection (FCD) Module}
\label{sec:dolos:mcdense}

The FCD module leverages graphs built over common activities performed by user accounts, in order to identify communities, each controlled by a different fraudster. Previous work has used graph based approaches to detect fraudulent behaviors, e.g., ~\cite{YL15,WGF17,SHYSLC17,hooi2016fraudar,XZ14}. Ye and Akoglu~\cite{YL15} quantified the chance of a product to be a spam campaign target, then clustered spammers on a 2-hop subgraph induced by the products with the highest chance values. Wang et. al~\cite{WGF17} leaveraged a novel Markov Random Field to detect fraudsters in social networks via guilt-by-association on directed graphs. Shen et al~\cite{SHYSLC17} introduced ``k-triangles'' to measure the tenuity of account groups and proposed algorithms to approximate the Minimum k-Triangle Disconnected Group problem. Hooi et al.~\cite{hooi2016fraudar} have shown that fraudsters have evolved to hide their traces, by adding spurious reviews to popular items. They introduced a class of ``suspiciousness'' metrics that apply to bipartite user-to-item graphs, and developed a greedy algorithm to find the subgraph with the highest suspiciousness metric.



\begin{figure}
\centering
\includegraphics[width=0.29\textwidth]{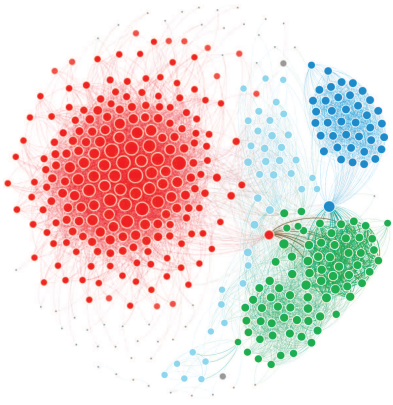}
\vspace{-5pt}
\caption{
{\bf Co-activity graph} of user accounts reviewing a popular horoscope app in Google Play (name hidden for privacy). Nodes are accounts. 4 Upwork workers each revealed to control the accounts of the same color.  Two accounts are connected if they post activities for similar sets of apps. Node sizes are a function of the account connectivity.}
\vspace{-15pt}
\label{fig:fraud}
\end{figure}

In contrast, the FCD module needs to solve the more complex problem of accurately identifying groups of user accounts such that each group is controlled by a {\it different} fraudster. In order to achieve this, we leverage the adversary Trait 4, that the accounts controlled by one fraudster are likely to have reviewed significantly more products in common than with the accounts controlled by another fraudster. We introduce {\it MCDense}, an algorithm that takes as input the \textbf{co-activity graph} of a product $A$, and outputs its {\it fraud components}, sets of user accounts, each potentially controlled by a different worker. We define the \textbf{co-activity graph} of a product $A$ as $G=(\mathcal{U},\mathcal{E}_{w})$, with a node for each user account that posted an activity for $A$ (see Figure~\ref{fig:fraud} for an illustration). Two nodes $u_i, u_j \in \mathcal{U}$ are connected by a weighted edge $e(u_i,u_j,w_{ij}) \in \mathcal{E}^w$, where the weight $w_{ij}$ is the number of products on which $u_i$ and $u_j$ posted activities in common.


		
		
			

MCDense, see Algorithm~\ref{alg:mcdense}, detects densely connected subgraphs, each subgraph being minimally connected to the other subgraphs. Given a graph $G=(\mathcal{U},\mathcal{E}_{w})$, its triangle density is $\rho(G)=\frac{t(V)}{{|V| \choose 3}}$, where $t(V)$ is the number of triangles formed by the edges in $\mathcal{E}^w$.

%

MCDense recursively divides the co-activity graph into two minimally connected subgraphs: the sum of the weights of the edges crossing the two subgraphs, is minimized.  If both subgraphs are more densely connected than the original graph (line 4) and the density of the original graph is below a threshold $\tau$, MCDense treats each subgraph as being controlled by different workers: it calls itself recursively for each subgraph (lines 5 and 6).  Otherwise, MCDense considers the undivided graph to be controlled by a single worker, and adds it to the set of identified components (line 8).

We have used the gold standard set of accounts controlled by the 23 fraudsters detailed in the previous section, to empirically set the $\tau$ threshold to 0.5, as the lowest density of the 23 groups of accounts revealed by the fraudsters was just above 0.5.

\noindent
{\bf MCDense converges and has $O(|\mathcal{E}_{w}| |\mathcal{U}|^3 )$ complexity}. To see that this is the case, we observe that at each step, MCDense either stops or, at the worst, ``shaves'' one node from $G$. The complexity follows then based on Karger's min-cut algorithm complexity~\cite{K93}.


\subsection{Component Attribution (CA) Module}
\label{sec:dolos:ca}

\begin{figure}[t]
\renewcommand{\baselinestretch}{0.1}
\begin{minipage}{0.45\textwidth}
\begin{algorithm}[H]
\begin{tabbing}
XX\=XX\=XX\=XX\=XX\=X\= \kill

{\bf Input}: G = $(\mathcal{U},\mathcal{E}_{w})$: input graph\\
\qquad \ \ \ \ \ n := $|\mathcal{U}|$\\
{\bf Output}: $\mathcal{C}$ := $\emptyset$: set of node components\\

1.\> MCDense(G)\{\\
2.\>\> \mbox{\bf{if}}\ (nodeCount(G) $<$ $\eta$)\ return;\\
3.\>\> ($G_1$, $G_2$) := weightMinCut(G);\\
4.\>\> \mbox{\bf{if}}\ (($\rho(G_1) > \rho(G)\ \&\ \rho(G_2) > \rho(G)$)\\
\>\>\> $\&$ ($\rho(G)$ $<$ $\tau$))\{\\
5.\>\>\> MCDense($G_1$); MCDense($G_2$);\\
6.\>\> \mbox{\bf{else}}\\
7.\>\>\> $\mathcal{C}$ := $\mathcal{C}$ $\cup$ G;\\
8.\>\>\> return;\\
9.\>\>\mbox{\bf{end if}}
\vspace{-5pt}
\end{tabbing}
\caption{MCDense: Min-Cut based Dense component detection. We set $\eta$ to 5 and $tau$ to 0.5.}
\label{alg:mcdense}
\end{algorithm}
\end{minipage}
\normalsize
\vspace{-15pt}
\end{figure}

Given a set of fraud worker profiles $\mathcal{FW}$ and a set of fraud components returned by the FCD module for a product $A$, the component attribution module identifies the workers likely to control the accounts in each component. To achieve this, {\scshape Dolos} leverages the unique writing style of human fraudsters to fuse elements from computational linguistics, e.g., ~\cite{OCCH11,LLKXXL11}, and author de-anonymization, e.g., ~\cite{OG16}. Specifically, we propose the following 2-step component attribution process:

\noindent
{\bf CA Training}.
Identify the products reviewed by the accounts controlled by each fraudster $W \in \mathcal{FW}$. For each such product, create a {\it review instance} that consist of all the reviews written by the accounts controlled by $W$ for $A$. Thus, each review instance contains only (but all) the reviews written from the accounts controlled by a single fraudster, for a single product. Extract stylometry features from each review instance of each fraudster, including character count, average number of characters per word, and frequencies of letters, uppercase letters, special characters, punctuation marks, digits, numbers, top letter digrams, trigrams, part of speech (POS) tags, POS digrams, POS trigrams, word digrams, word trigrams and of misspelled words. Train a supervised learning algorithm on these features, that associates the feature values of each review instance to the fraudster who created it.

\noindent
{\bf Attribution}.
Let $\mathcal{C}$ denote the set of components returned by MCDense for a product $A$.  For each component $C \in \mathcal{C}$, group all the reviews written by the accounts in $C$ for product $A$, into a review instance, $r$. Extract $r$'s stylometry features and use the trained classifier to determine the probability that $r$ was authored by each of the fraudsters in $\mathcal{FW}$. Output the identity of the fraudster with the highest probability of having authored $r$. 


\section{Empirical Evaluation}
\label{sec:evaluation}

\begin{figure*}
\centering
\subfigure[]
{\label{fig:cdf:mcdense:numofcomp}{\includegraphics[width=0.32\textwidth]{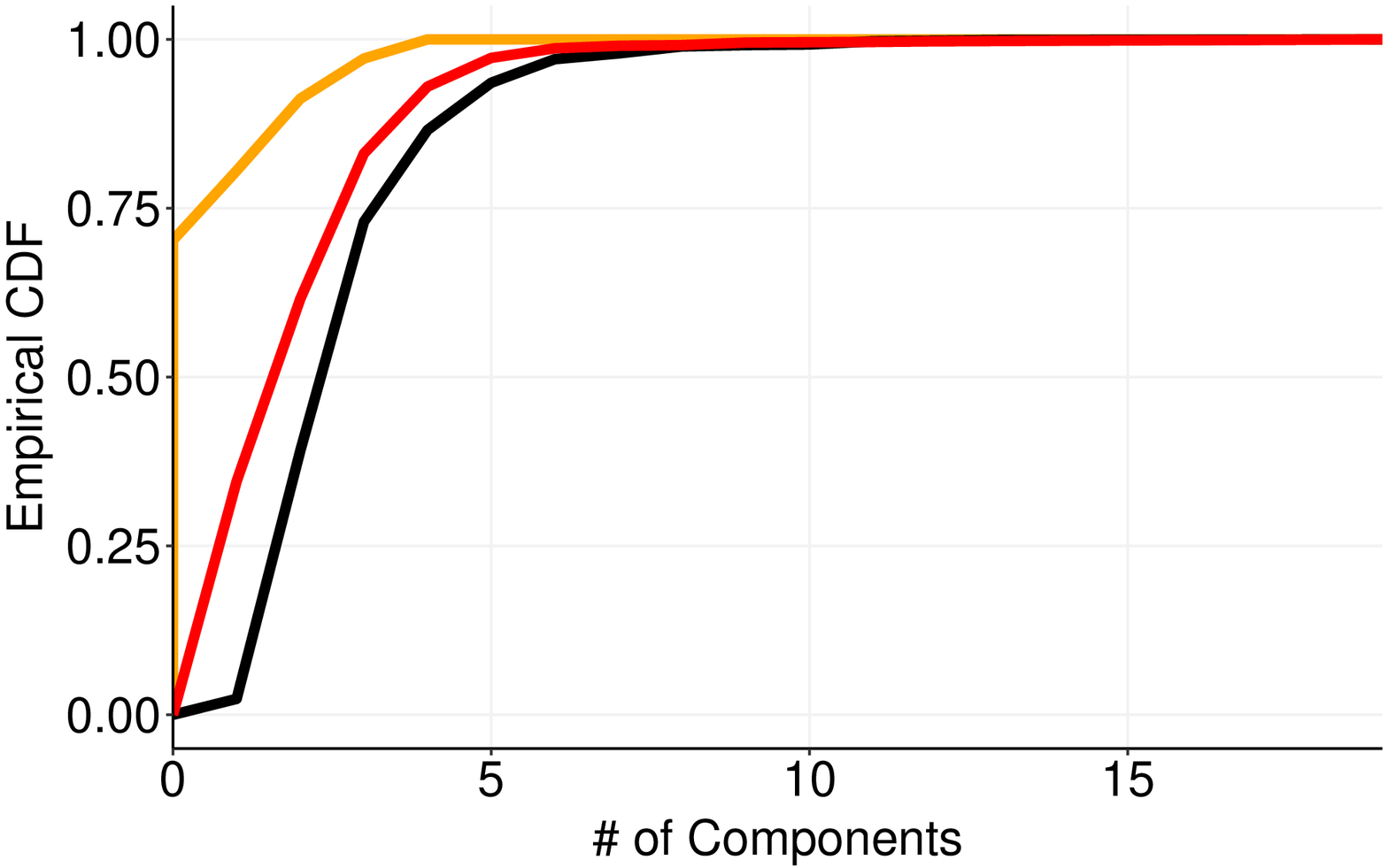}}}
\subfigure[]
{\label{fig:cdf:mcdense:componentdensity}{\includegraphics[width=0.32\textwidth]{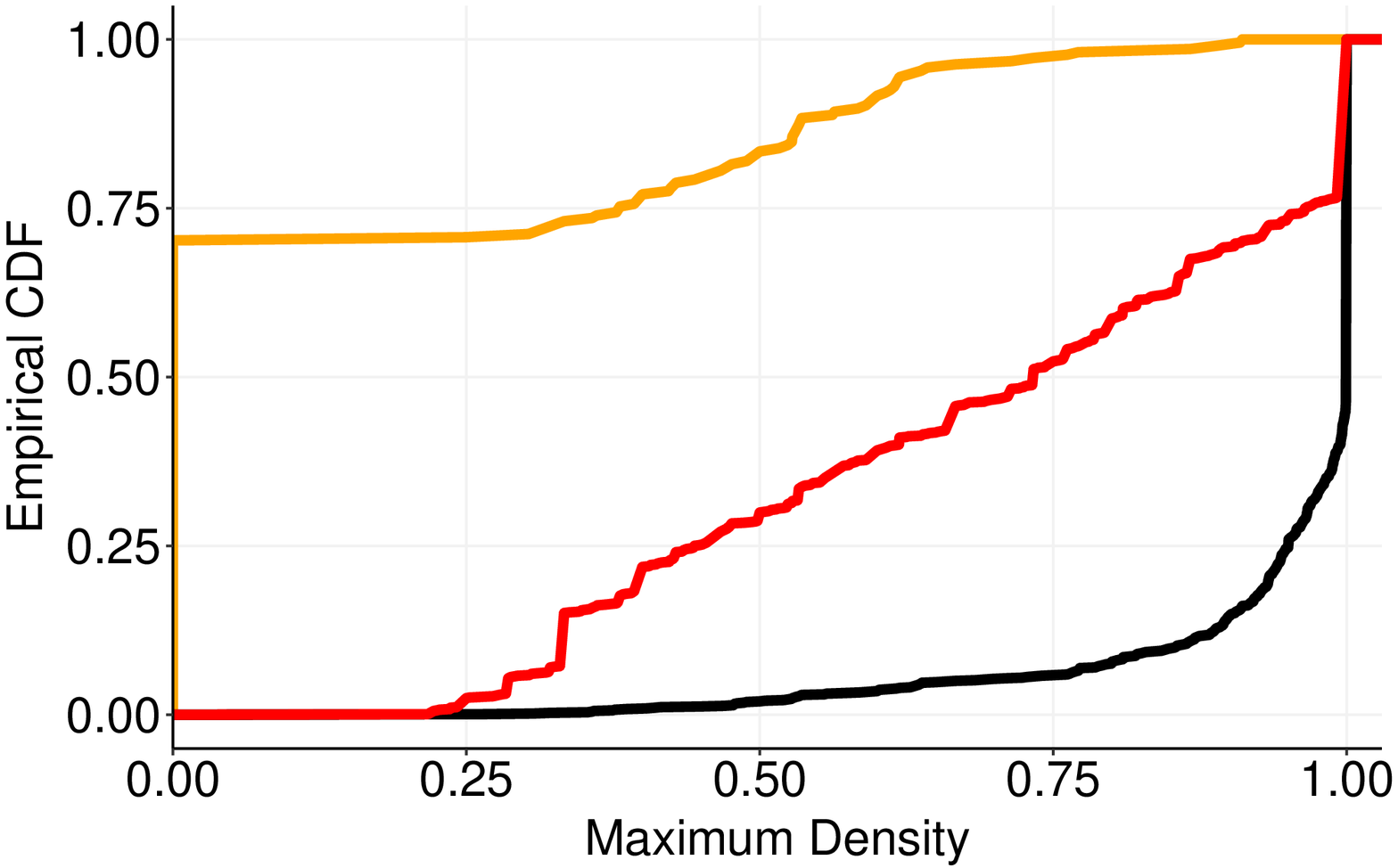}}}
\subfigure[]
{\label{fig:cdf:mcdense:maxcompsize}{\includegraphics[width=0.32\textwidth]{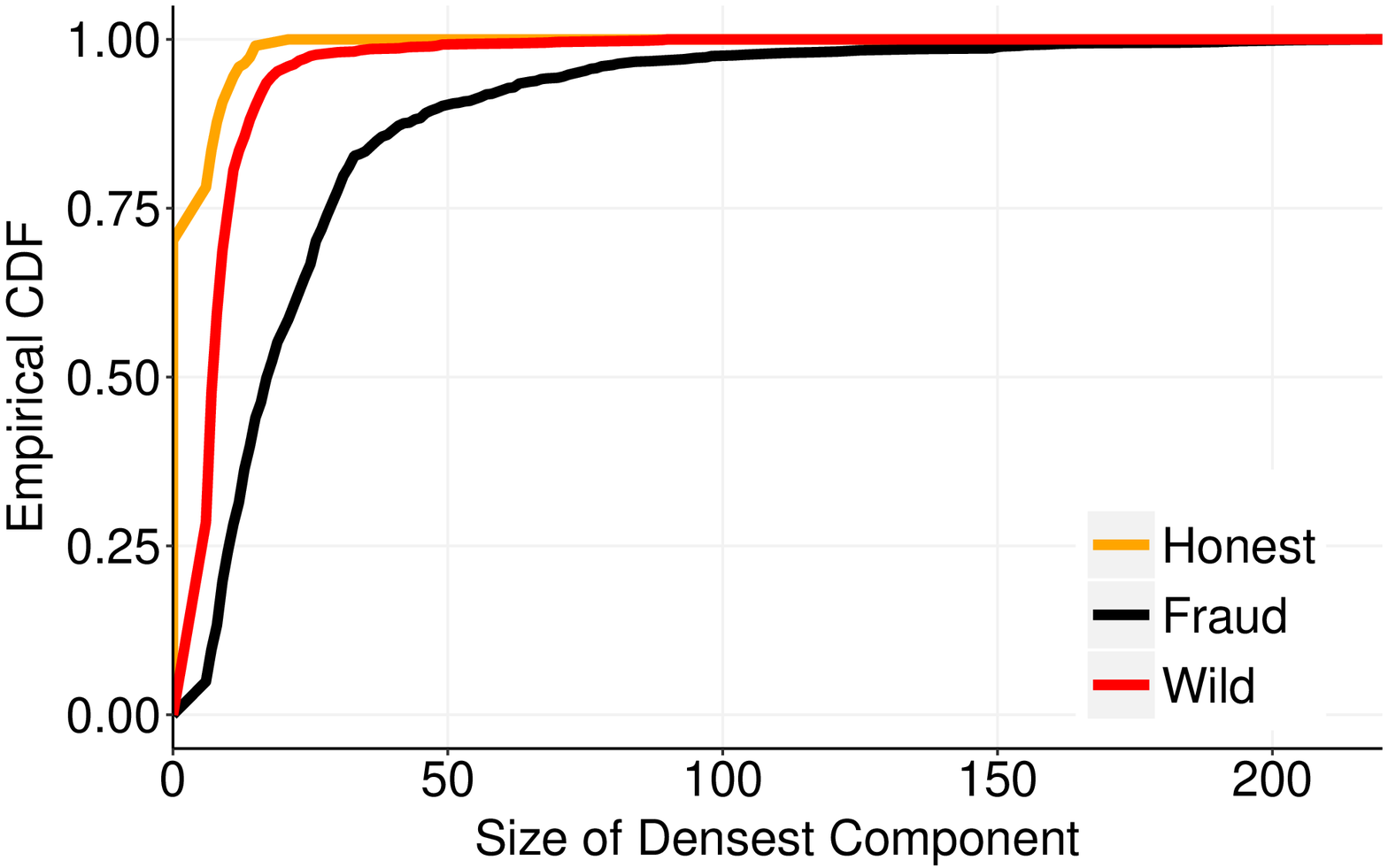}}}
\vspace{-15pt}
\caption{MCDense: Cummulative distribution function (CDF) over 640 fraud, 219 honest, and 1,056 suspicious ``wild'' apps, of per-app (a) number of components  of at least 5 accounts, (b) maximum density of an identified component and (c) size of densest component.
We observe significant differences between fraud and honest apps.}
\label{fig:cdf}
\end{figure*}

In this section we compare the results of {\scshape Dolos} on fraud and honest apps, evaluate its de-anonymization accuracy, and present its results on 13,087 apps. Further, we compare MCDense with DSG, an adapted dense sub-graph detection solution.

\subsection{Fraud vs. Honest Apps}
\label{sec:evaluation:honestfraud}

We evaluate the ability of {\scshape Dolos} to discern differences between fraudulent and honest apps. For this, we have first selected 925 candidate apps from the longitudinal app set, that have been developed by Google designated ``top developers''. We have filtered the apps flagged by VirusTotal. We have manually investigated the remaining apps, and selected a set of 219 apps that (i) have more than 10 reviews and (ii) were developed by reputable media outlets (e.g., Google, PBS, Yahoo, Expedia, NBC) or have an associated business model (e.g., fitness trackers). We have collected 38,224 reviews and their associate user accounts from these apps.

Figure~\ref{fig:cdf:mcdense:numofcomp} compares the CDF of the number of components (of at least 5 accounts) found by MCDense per each of the 640 fraud apps vs. the 219 honest apps. MCDense found that all the fraud apps had at least 1 component, however, 70\% of the honest apps had no component. The maximum number of components found for fraud apps is 19 vs. 4 for honest apps. Figure~\ref{fig:cdf:mcdense:componentdensity} compares the CDF of the maximum edge density (ratio of number of edges to maximum number of edges possible) of a component identified by MCDense per fraud vs. honest apps. 94.4\% of fraud apps have density more than 75\% while only 30\% of the honest apps have a cluster with density larger than 0. The increase is slow, with 90\% of the honest apps having clusters with density of 60\% or below. Figure~\ref{fig:cdf:mcdense:maxcompsize} compares the CDF of the size of the per-app densest component found for fraud vs. honest apps. 80\% of the fraud apps vs. only 7\% of the honest apps, have a densest component with more than 10 nodes. The largest, densest component has 220 accounts for a fraud app, and 21 accounts for an honest app. We have manually analyzed the largest, densest components found by MCDense for the honest apps and found that they occur for users who review popular apps such as the Google, Yahoo or Facebook clients, and users who share interests in, e.g., social apps or games. 


\subsection{De-Anonymization Performance}
\label{sec:evaluation:f2f}

\begin{figure*}
\centering
\subfigure[]
{\label{fig:worker:instances}{ \includegraphics[width=0.47\textwidth]{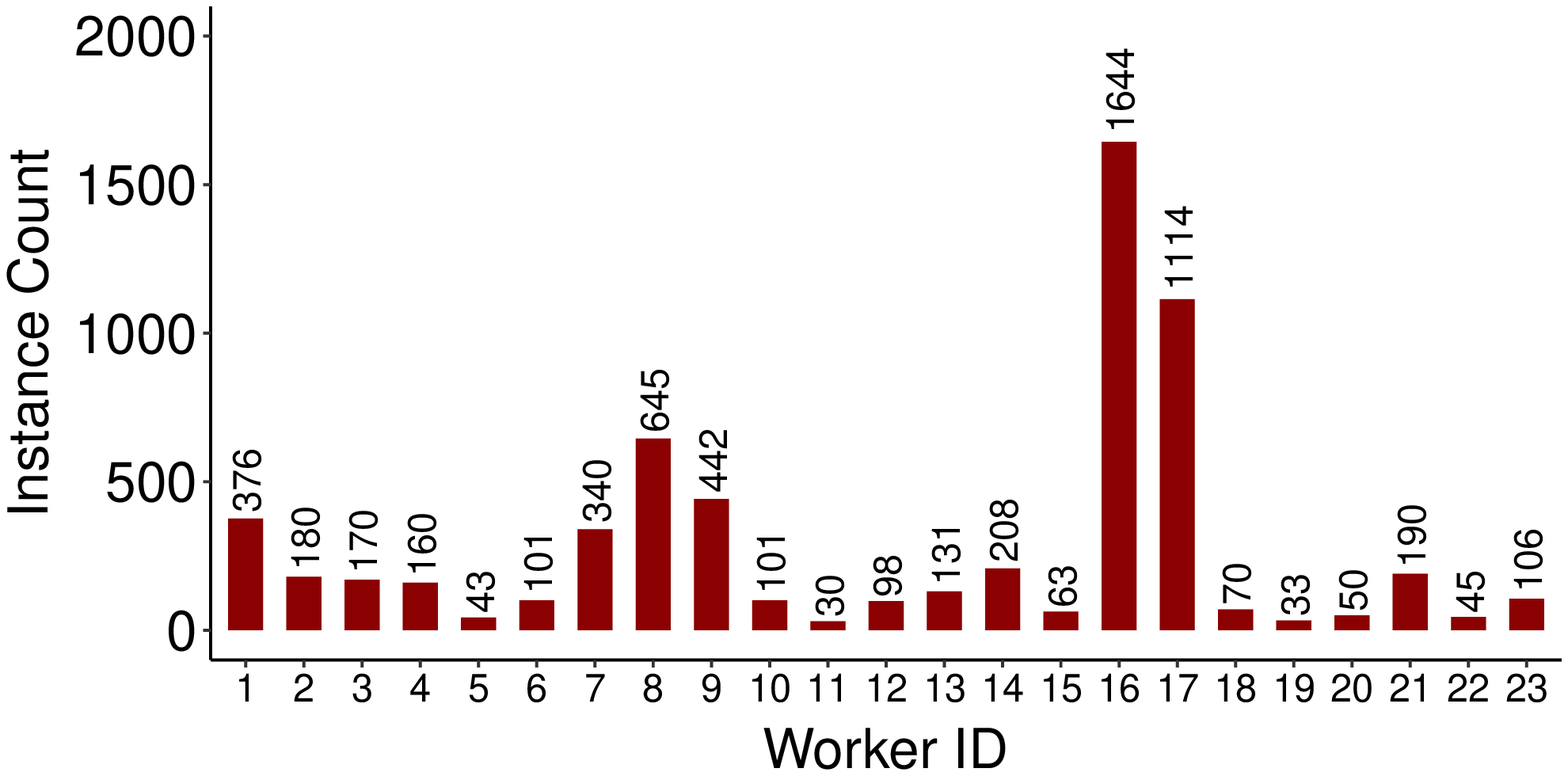}}}
\subfigure[]
{\label{fig:worker:break}{
\includegraphics[width=0.45\textwidth]{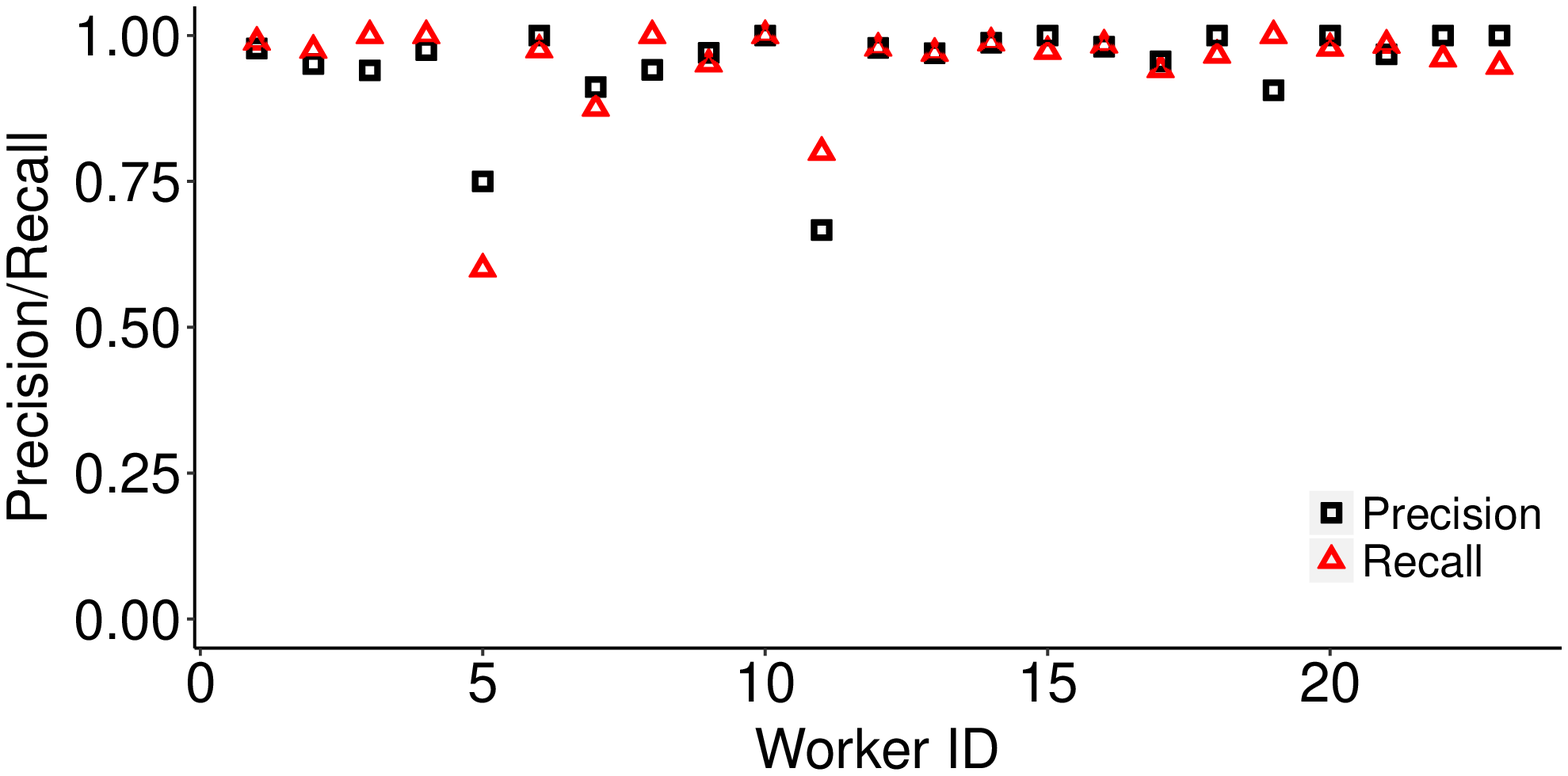}}}
\vspace{-15pt}
\caption{(a) Number of review instances collected from each of the 23 fraudster. Each review instance has at least 5 reviews, written by the accounts controlled by a single fraudster, for a single app.
(b) {\scshape Dolos} per-worker attribution precision and recall, over the 1,690 review instances of 640 fraud apps, exceed 87\% for 21 out of the 23 fraudsters.}
\vspace{-5pt}
\label{fig:worker:instances2}
\end{figure*}

We have implemented the CA module using a combination of JStylo~\cite{JStylo} and supervised learning algorithms. We have collected the 111,714 reviews posted from the 2,664 attributed, fraudster controlled user accounts of $\S$~\ref{sec:model:fpc}. The reviews were posted for 2,175 apps. We have grouped these reviews into instances, and we have filtered out those with less than 5 reviews.
Figure~\ref{fig:worker:instances} shows their distribution  among the 23 fraudsters who authored them.


We have evaluated the performance of {\scshape Dolos} (MCDense + CA) using a leave-one-out cross validation process over the 640 fraud apps (and their 1,690 review instances). We have used several supervised learning algorithms, including k-nearest neighbors (k-NN), Random Forest (RF), Decision Trees (DT), Naive Bayes (NB), and Support Vector Machine (SVM). In each experiment we report the top 3 performers.




\begin{table}
\setlength{\tabcolsep}{.16677em}
\centering
\textsf{
\begin{tabular}{l | r r | r r}
\toprule
\textbf{Algo} & \textbf{Top 1} & \textbf{(TPR)} & \textbf{Top 3} & \textbf{Top 5}\\
\midrule
k-NN (IBK) & \textbf{1608} & (\textbf{95.0\%}) & \textbf{1645} & 1646\\
RF {(\small Random Forest)} & 1487 & (87.9\%) & 1625 & \textbf{1673}\\
DT {(\small Decision Tree)} & 1126 & (66.5\%) & 1391 & 1455\\
\bottomrule
\end{tabular}
}
\caption{{\scshape Dolos} attribution performance for the 1,690 instances of the 640 fraud apps. k-NN achieved the best performance: It correctly identifies the workers responsible for 95\% (1608) of the instances.
}
\label{table:instances}
\vspace{-15pt}
\end{table}

\noindent
{\bf Instance level performance}.
Table~\ref{table:instances} shows the number of instances correctly attributed by {\scshape Dolos} (out of the 1,690 instances of the 640 fraud apps) and corresponding true positive rate, as well as the number of instances where the correct worker is among {\scshape Dolos}' top 3 and top 5 options. k-NN achieved the best performance, correctly identifying the workers responsible for posting 95\% of the instances. We observe that k-NN correctly predicts the authors of 95\% of the instances. Figure~\ref{fig:worker:break} zooms into per-fraudster precision and recall, showing the ability of {\scshape Dolos} to identify the instances and only the instances of each of the 23 workers. For $21$ out of $23$ workers, the {\scshape Dolos} precision and recall both exceed $87$\%.


\noindent
{\bf App level performance}.
Table~\ref{table:f2f:apps:recall} shows that when using k-NN, {\scshape Dolos} correctly identified at least 1 worker per app, for $97.5$\% of the fraud apps, and identified at least $90$\% of the workers in each of $87$\% of the fraud apps. Table~\ref{table:f2f:apps:precision} shows that the precision of {\scshape Dolos} in identifying an app's workers exceeds $90$\% for $69$\% of the apps.

\begin{table}[t]
\setlength{\tabcolsep}{.16677em}
\centering
\small
\textsf{
\begin{tabular}{l | r | r | r | r}
\toprule
\textbf{Algo} & \textbf{1 worker} & \textbf{50\%-recall} & \textbf{70\%-recall} & \textbf{90\%-recall}\\
\midrule
RF & 624 & 622 & 537 & 465\\
SVM & 574 & 517 & 325 & 284\\
k-NN & \textbf{625} & {\bf 625} & {\bf 585} & {\bf 557}\\
\bottomrule
\end{tabular}}
\caption{{\scshape Dolos} app level recall: the number of apps for which {\scshape Dolos} has a recall value of at least 50\%, 70\% and 90\%. k-NN identifies at least one worker for 97.5\% of the 640 fraud apps, and 90\% of the workers of each of 557 (87\%) of the apps.
\label{table:f2f:apps:recall}}
\vspace{-5pt}
\end{table}

\begin{table}
\setlength{\tabcolsep}{.16677em}
\centering
\small
\textsf{
\begin{tabular}{l | r | r | r}
\toprule
\textbf{Algo} & \textbf{50\%-prec} & \textbf{70\%-prec} & \textbf{90\%-prec}\\
\midrule
RF & 573 & 434 & 359\\
SVM & 460 & 249 & 209\\
k-NN & \textbf{578} & \textbf{483} & \textbf{444}\\
\bottomrule
\end{tabular}}
\caption{App level precision: the number of apps where its precision is at least 50\%, 70\% and 90\%. The precision of {\scshape Dolos} when using k-NN exceeds 90\% for 69\% of the fraud apps.}
\label{table:f2f:apps:precision}
\vspace{-15pt}
\end{table}


\noindent
{\bf Developer tailored search rank fraud}.
Upon closer inspection of the {\scshape Dolos} identified clusters, we found numerous cases of clusters consisting of user accounts who reviewed almost exclusively apps created by a single developer.
%
%
We conjecture that those user accounts were created with the specific goal to review the apps of the developer, e.g., by the developer or their employees.


\subsection{{\scshape Dolos} in the Wild}

To understand how Dolos will perform in real life, we have randomly selected 13,087 apps from Google Play, developed by 9,430 distinct developers. We monitored these apps over more than 6 months, and recorded their changes once every 2 days. This enabled us to collect up to 7,688 reviews per app, exceeding Google's one shot limit of 4,000 reviews. We collected the data of the 586,381 distinct reviewers of these apps, and built their co-activity graphs.

MCDense found at least 1 dense component of at least 5 accounts in 1,056 of the 13,087 apps (8\%). Figure~\ref{fig:cdf} compares the results of MCDense on the 1,056 apps, with those for the fraud and honest apps. The CDF of the number of components found by MCDense for these ``wild'' apps is closer to that of the fraud apps than to the honest apps: up to 19 components per app, see Figure~\ref{fig:cdf:mcdense:numofcomp}. The CDF of the maximum density of per app components reveals that 231 of the 1,056 apps (or 21.87\%) had at least 1 component with edge density 1 (complete sub-graphs). The CDF of the size of the densest components (Figure~\ref{fig:cdf:mcdense:maxcompsize}) found per each of the wild apps shows that similar to the 640 fraud apps, few of these apps have only 0 size densest components. The largest component found by MCDense for these apps has 90 accounts.

\noindent
{\bf Validation of fraud suspicions}.
Upon close inspection of the 231 apps that had at least 1 component with edge density of 1 (i.e., clique), we found the following further evidence of suspicious fraud being perpetrated. (1) {\bf Targeted by known fraudsters}: 169 of the 231 apps had received reviews from the 23 known fraudsters ($\S$~\ref{sec:model:fpc}). One app had received reviews from 10 of the fraudsters. (2) {\bf Review duplicates}: 223 out of the 231 apps have received 10,563 {\it duplicate} reviews (that replicate the text of reviews posted for the same app, from a different account), or 25.55\% of their total 41,339 reviews. One app alone has 1,274 duplicate reviews, out of a total of 4,251 reviews. (3) {\bf Fraud re-posters}: our longitudinal monitoring of apps enabled us to detect fraud re-posters, accounts who re-post their reviews, hours to days after Google Play filters them out. One of the 231 apps received 37 fraud re-posts, from the same user account.

\section{MCDense Evaluation}
\label{sec:evaluation:fcd}



%
%


	

We compare MCDense against DSG, a densest subgraph approach that we adapt based on~\cite{T15}. DSG iteratively identifies multiple dense subgraphs of an app's co-activity graph $G = (U, E)$, each suspected to belong to a different worker. DSG peels off nodes of $G$ until it runs out of nodes. During each ``peeling'' step, it removes the node that is least connected to the other nodes. After removing the node, the algorithm computes and saves the density of the resulting subgraph. The algorithm returns the subgraph with the highest density. We use the ``triangle'' density definition proposed in~\cite{T15}, $\rho_D = \frac{t(U)}{|U|}$, where $t(U)$ is the number of triangles formed by the vertices in $U$.  DSG uses this greedy strategy iteratively: once it finds the densest subgraph $D$ of $G$, DSG repeats the process, to find the densest subgraph in $G - D$.

\begin{figure}[t!]
\centering
\includegraphics[width=2.9in]{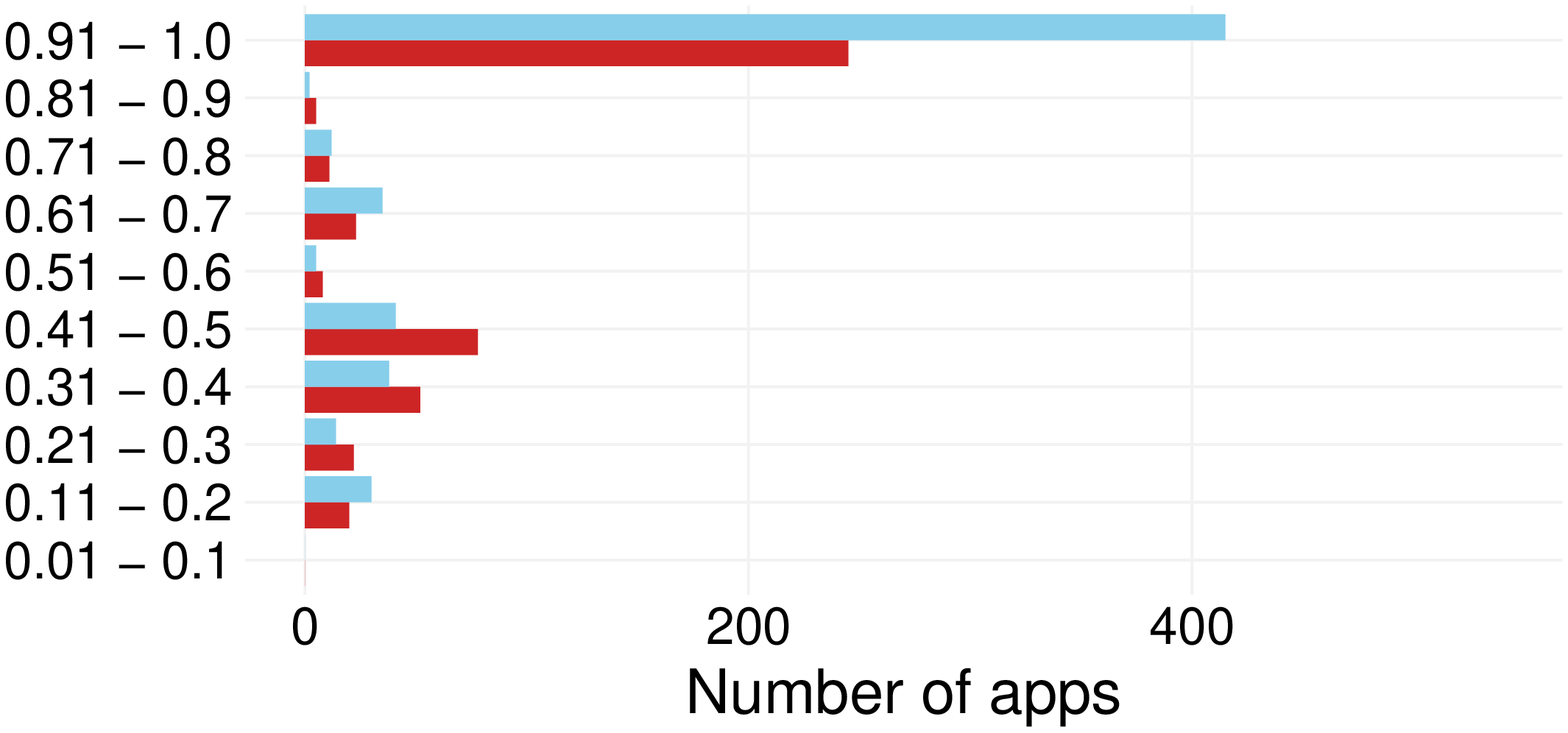}
\vspace{-5pt}
\caption{Comparison of the distribution of coverage scores for MCDense and DSG over 640 fraud apps. The $y$ axis shows the $p_1$ value, and the $x$ axis shows the number of apps for which MCDense and DSG achieve that $p_1$ value, when $p_2$ is 90\%. {\bf MCDense outperforms DSG} providing (90\%+, 90\%)-coverage for 416 (65\%) of the apps vs. DSG's 245 apps.}
\label{fig:pcoverage:90}
\vspace{-5pt}
\end{figure}

To compare MCDense and DSG, we introduce a coverage score.
%
Let $\mathcal{C} = \{C_1, .., C_c, H_C\}$ be the partition of 
of the user accounts who reviewed an app $A$,
returned by a fraud-component detection algorithm: $\forall a_i, a_j \in C_l$, are considered to be controlled by the same worker, and $H_C$ is the set of accounts considered to be honest. To quantify how well the partition $\mathcal{C}$ has detected the worker accounts $\overline{W_1}, .., \overline{W_w}$ who targeted $A$, we propose the ``coverage'' measure of worker $W_i \in S$ by a partition $C$, as $cov_{i}(C) = \frac{|\overline{W_i} \cap (C_1 \cup .. \cup C_c)|}{|\overline{W_i}|}$.  Given $p \in [0,1]$, we say that $W_i$ is ``$p$-covered'' by $C$ if $cov_{i}(C) \ge p$. Then, we say that partition $C$ provides a $(p_1, p_2)$-coverage of the worker set $S$, if $p_1$ percent of the workers in $S$ are $p_2$-covered by $C$.

Figure~\ref{fig:pcoverage:90} compares the distribution of the coverage scores for MCDense and DSG over the $640$ fraud apps. It shows that the number of apps for which at least 90\% of their workers are at least 90\%-covered is twice as high for MCDense than for DSG.

\section{Conclusions}
\label{sec:conclusions}

We introduced the fraud de-anonymization problem for search rank fraud in online services. We have collected fraud data from crowdsourcing sites and the Google Play store, and we have performed a user study with crowdsourcing fraudsters. We have proposed {\scshape Dolos}, a fraud de-anonymization system. {\scshape Dolos} correctly attributed $95$\% of the fraud detected for $640$ Google Play apps, and identified at least $90$\% of the workers who promoted each of $87$\% of these apps. {\scshape Dolos} identified 1,056 out of 13,087 monitored Google Play apps, to have suspicious reviewer groups, and revealed a suite of observed fraud behaviors.

\section{Acknowledgments}
\label{sec:acknowledgments}

This research was supported by NSF grants CNS-1527153, CNS-1526254 and CNS-1526494, and by the Florida Center for Cybersecurity.

\bibliographystyle{ACM-Reference-Format}
\bibliography{attribution,bogdan,crowdsource,graph,lbp,ml,reviews,social.fraud,stylo}

\end{document}